\documentclass{article}

\usepackage{arxiv}

\usepackage[utf8]{inputenc} 
\usepackage[T1]{fontenc}    
\usepackage{hyperref}       
\usepackage{url}            
\usepackage{booktabs}       
\usepackage{amsfonts}       
\usepackage{nicefrac}       
\usepackage{microtype}      
\usepackage{lipsum}		
\usepackage{graphicx}
\usepackage[numbers, sort&compress]{natbib}
\usepackage{doi}
\usepackage{amssymb}
\usepackage{amsmath}
\graphicspath{ {./img/} }

\makeatletter
\renewcommand{\@toptitlebar}{}      
\renewcommand{\@bottomtitlebar}{}   
\makeatother

\title{Geometry and Kinematics of Molecular Cloud Substructures in the Second Galactic Quadrant$^\ast$}

\author{
  Wen Ge$^{1,2}$ \quad Fujun Du$^{1,2}$$^\dagger$ \\
  $^1$ Purple Mountain Observatory and Key Laboratory of Radio Astronomy, Chinese Academy of Sciences, \\
  10 Yuanhua Road, Qixia District, Nanjing 210023, People's Republic of China\\
  $^2$ School of Astronomy and Space Science, University of Science and Technology of China, Hefei 230026,\\
  People's Republic of China
}

\date{}
\renewcommand{\headeright}{}
\renewcommand{\undertitle}{}
\renewcommand{\shorttitle}{Wen Ge et al.: Geometry and Kinematics of Molecular Cloud Substructures}

\begin{document}
\maketitle

\renewcommand{\thefootnote}{}   
\setcounter{footnote}{0}        

\footnotetext{\small \hspace*{0.8em}Received 2026-04-01, revised version 2026-05-20. Accepted 2026-05-29.\\
\small \hspace*{2.4em}$^\ast$This work is supported by National Key R \& D Program of China with grant 2023YFA1608000 and National SKA Program of China with grant 2025SKA0140100.\\
\small \hspace*{2.4em}$^\dagger$fjdu@pmo.ac.cn}

\begin{abstract}
We analyze the geometry and kinematics of substructures within molecular clouds identified in an unbiased catalog from the MWISP (Milky Way Imaging Scroll Painting) survey. These substructures are defined as spatially connected regions enclosed by the 20\% peak-integrated-intensity contour of each cloud. After applying selection criteria on voxel size and excluding structures truncated by map boundaries, we construct a sample and quantify their projected morphology using the projected scale ratio $R=\Delta b/(\Delta l\cdot\cos b)$ based on interpolated integrated-intensity contours. This ratio essentially measures $\tan\theta$ where $\theta$ is the plane-of-sky angle of an elongated filament relative to the Galactic plane.
The resulting sample exhibits a median $R=0.96$, indicating a slight but systematic preference for elongation along Galactic longitude. This tendency becomes more pronounced at larger spatial scales.
We further investigate the relative orientations among the structural major axes, velocity-gradient directions, and plane-of-sky magnetic-field orientations derived from Planck data for a subsample of well-defined structures. 
We find that, for cloud structures within our sample, with physical scale $\sim 0.3$ to $\sim 30\,\mathrm{pc}$, velocity gradients tend to be perpendicular to the structural major axes, while magnetic-field orientations are generally aligned parallel to them. This scale range differs from those typically probed in studies of dense cores ($\sim 0.05\,\mathrm{pc}$) and GMC-scale structures ($\gtrsim 10$--$100\,\mathrm{pc}$), which have reported scale-dependent variations in relative orientations. In addition, the alignment between velocity gradients and magnetic fields shows a gradual weakening with increasing physical scale.
These results suggest that the observed anisotropy of molecular cloud substructures may arise from a combination of large-scale Galactic dynamics, anisotropic gas motions, and magnetic fields, with the relative importance of these effects varying with scale.
\end{abstract}


\section{Introduction}
\label{sec:intro}
The study of molecular cloud morphology is instrumental in interstellar medium research, as it provides key insights into the origin and evolution of these clouds \cite{Hacar+etal+2023}.
The spatial distribution and geometric properties of molecular clouds reflect the interplay of turbulence, gravity, magnetic fields, and large-scale Galactic dynamics \cite{Heyer+etal+2015}.
Understanding how these processes shape the observed cloud structures is essential for uncovering the mechanisms that control molecular cloud formation, fragmentation, and the subsequent onset of star formation \cite{McKee+etal+2007, Andre+etal+2014}.

Molecular clouds often exhibit complex hierarchical structures \cite{Heyer+etal+2015}.
In some studies, they are described as fractal structures, where the fractal dimension serves as a quantitative measure of their boundaries \cite{Falgarone+etal+1991, Stutzki+etal+1998, Beattie+etal+2019, Yan+etal+2025}.
The definition of molecular cloud boundaries themselves is not unique, and may depend on observational details \cite{Yan+etal+2022}.

The aspect ratio is one of the parameters commonly used in studies of molecular cloud morphology. In most cases, the aspect ratio is defined as the ratio between the length and width of a molecular cloud. For instance, the ``Nessie" infrared dark cloud identified by Jackson et al.\cite{Jackson+etal+2010} has a length of about $80\,\mathrm{pc}$ and exhibits an extremely large aspect ratio of up to 150.
Goodman et al.\cite{Goodman+etal+2014} further suggested that this cloud may have an even larger extent and aspect ratio.
Filament aspect ratios have been extensively studied in filamentary molecular clouds. Using \textit{Herschel} observations, Arzoumanian et al.\cite{Arzoumanian+etal+2019} identified nearby filaments with typical aspect ratios of about 5 and extreme values above 30. Studies of giant molecular filaments and Galactic ``bones'' reported substantially larger aspect ratios, reaching $\gtrsim 100$ in some cases \cite{Zucker+etal+2015,Wang+etal+2016}. These works mainly focused on identifying highly elongated filamentary structures. In contrast, the projected scale ratio used in this work is not intended to define filamentary clouds or infer intrinsic elongations, but rather to characterize the projected directional anisotropy of molecular-cloud substructures in Galactic coordinates.
In this work, instead of the intrinsic length-width definition, we use a projected scale ratio, defined as the ratio between the angular extents in Galactic latitude ($\Delta b$) and Galactic longitude ($\Delta l$), to quantify directional anisotropy in Galactic coordinates. For an elongated structure, this ratio can be interpreted approximately as $\tan\theta$, where $\theta$ is the plane-of-sky orientation relative to the Galactic plane (see Section~\ref{SubSec:Selection_and_PSR_Definition} for details).

The velocity gradient is another fundamental parameter to study the kinematics of molecular clouds.
Previous studies have typically characterized large-scale velocity gradients in molecular clouds by fitting a plane to the velocity field or by assuming a global linear gradient \cite{Goodman+etal+1993, Koda+etal+2006, Imara+etal+2011}. While these methods provide a convenient description of global kinematics, they implicitly assume a coherent, large-scale linear velocity field and do not explicitly account for spatial variations in gradient directions.
In this work, we complement these approaches by statistically characterizing the distribution of local velocity-gradient directions. This allows us to define both a principal gradient orientation and a quantitative measure of directional coherence, and to investigate the relative orientations among cloud morphology, gas kinematics, and magnetic fields using a common sample and a consistent statistical methodology.

Magnetic fields can play a key role in shaping cloud structures and regulating gas motions \cite{Crutcher+etal+2012} during the formation and evolution of molecular clouds. Several studies have explored the relationship between magnetic field orientations and the alignment of structures within molecular clouds. For example, Li et al.\cite{Li+etal+2011} reported that magnetic fields are aligned with the spiral arms in M33, and more recently, Sun et al.\cite{Sun+etal+2024} found that striations are well aligned with the magnetic fields. These results suggest a close connection between cloud morphology, gas kinematics, and magnetic-field geometry.
Jiao et al.\cite{Jiao+etal+2024} found a change of the relative alignment between magnetic field and gas structure as the column density is increased.
We find that at the scales covered by the survey data we are using, the cloud structures tend to be aligned with magnetic fields.

We use an unbiased catalog of molecular clouds in the second Galactic quadrant from the MWISP (Milky Way Imaging Scroll Painting) survey to investigate their morphological characteristics and possible physical origins. In Section~\ref{Sec:Data_and_Method} we describe the MWISP data and analysis methods in detail. In Section~\ref{Sec:Results_and_Discussion} we present our results and related discussions, and Section~\ref{Sec:Conclusions} is a summary of the main findings.

\section{Data and Method} \label{Sec:Data_and_Method}

\subsection{The MWISP Project} \label{SubSec:The_MWISP_Project}

The MWISP project is a systematic survey of CO and its isotopologues on the Galactic plane with the $13.7\,\mathrm{m}$ single-dish telescope of the Purple Mountain Observatory \cite{Yang+etal+2026}. Three molecular spectral lines of $^{12}\mathrm{CO}/^{13}\mathrm{CO}/\mathrm{C}^{18}\mathrm{O}\;J=1-0$ are observed at the same time. The typical system temperatures are $\sim 250\,\mathrm{K}$ for $^{12}$CO and $\sim 140\,\mathrm{K}$ for $^{13}$CO and C$^{18}$O, respectively \cite{Su+etal+2019}. The data used in this work are in the form of a fits cube, where the first two dimensions represent the Galactic longitude and latitude, while the third dimension represents the local standard-of-rest velocity ($V_{\mathrm{LSR}}$). The angular resolution of the data is $50\,\mathrm{arcsec}$, and the velocity resolutions are $0.16\,\mathrm{km\cdot s}^{-1}$ for the $^{12}$CO lines and $0.17\,\mathrm{km\cdot s}^{-1}$ for the $^{13}$CO and C$^{18}$O lines. The typical rms noise levels are $\sim0.5\,\mathrm{K}$ for $^{12}$CO and $\sim0.3\,\mathrm{K}$ for $^{13}$CO and C$^{18}$O.

The molecular clouds that we use for this work are obtained from the $^{12}\mathrm{CO}(1 - 0)$ spectral line data using the DBSCAN algorithm \cite{Ester+etal+1996}, with Galactic longitudes $104.75\,^\circ \leqslant l \leqslant 150.25\,^\circ$, Galactic latitudes $\vert b\vert \leqslant 5.25\,^\circ$, and line-of-sight velocity of $-95\,\mathrm{km\cdot s}^{-1} \leqslant V_{\mathrm{LSR}} \leqslant 25\,\mathrm{km\cdot s}^{-1}$ \cite{Yuan+etal+2021}.
Details about the parameters of the DBSCAN algorithm applied to molecular clouds are described in Yan et al.\cite{Yan+etal+2021}.
The minimum cutoff on the PPV data cubes is set to 2$\sigma $ ($\sim 1\,\mathrm{K}$ for $^{12}$CO lines). The velocity information derived from the $^{12}$CO data primarily reflects the large-scale kinematics of the extended molecular gas envelope. The results presented in this work therefore provide a statistical characterization of velocity-structure relations on cloud scales and do not probe the internal dynamics of dense cores.

\subsection{Planck \texorpdfstring{$353\, \mathrm{GHz}$}{353 GHz} Observation} \label{SubSec:Planck_353_GHz_Observation}

The Planck all-sky survey \cite{Planck+etal+2011} measures linear polarization in seven frequency bands from $30\, \mathrm{GHz}$ to $353\, \mathrm{GHz}$, with the $353\, \mathrm{GHz}$ band providing the highest sensitivity for dust polarization \cite{Planck+etal+2015}. At this frequency, contamination from the cosmic microwave background is negligible when observing molecular clouds \cite{Soler+etal+2019}.
The $353\, \mathrm{GHz}$ polarization maps have an angular resolution of $4.8\,^\prime$ in the HEALPix format with $N_{\mathrm{side}} = 2048$, corresponding to a pixel size of about $1.7\,^\prime$.

We compute the polarization position angle ($\psi$) as
\begin{equation}
    \psi = -\frac{1}{2}\, \mathrm{arctan}(U/Q) \, ,
\end{equation}
where $\psi = 0\,^\circ$ points north and increases counterclockwise following the IAU convention \cite{Planck+etal+2016}. The plane-of-sky magnetic field orientation ($\theta_B$) is then obtained as
\begin{equation}
    \theta_B = \psi - 90\,^\circ \, .
\end{equation}

\subsection{Morphological Selection and Definition of the Projected Scale Ratio}
\label{SubSec:Selection_and_PSR_Definition}

For the current purpose, to study the morphology of molecular clouds, we first exclude molecular clouds with angular scales that are either too large or too small. The former may correspond to molecular cloud complexes, while the latter are affected by insufficient angular resolution, resulting in unclear internal structures. We also exclude molecular clouds with limited velocity extents, as their signal-to-noise ratios (S/N) may be too low for reliable identification. In practice, we adopt $3 \times 3 \times 3$ ($1.5\,\mathrm{arcmin} \times 1.5\,\mathrm{arcmin} \times 0.474\,\mathrm{km\cdot s}^{-1}$) and $40 \times 40 \times 30$ ($20\,\mathrm{arcmin} \times 20\,\mathrm{arcmin} \times 4.740\,\mathrm{km\cdot s}^{-1}$) as the strict lower and upper bounds for voxel sizes, in the sense that only clouds with voxel sizes strictly between them are included. We use angular-size selection not as a substitute for sample classification based on physical scale, but rather as a practical criterion to ensure the robustness of the derived physical quantities under the limited spatial resolution of the observational data. We also note that a small fraction of the molecular clouds lie on the spatial or velocity boundaries of the region under study.  These are excluded and we ultimately obtain a sample of 18,078 molecular clouds.

In this work, we define the $\Delta b/\Delta l$ ratio as
\begin{equation}
    \frac{\Delta b}{\Delta l \cdot \cos b} \, ,
\end{equation}
where $\Delta b$ and $\Delta l$ represent the maximum extents of the molecular cloud in Galactic latitude and longitude, respectively. The factor $\cos b$ corrects for the projection effect at Galactic latitude $b$; since the clouds in our sample lie at relatively low Galactic latitudes, this correction is minor ($\cos 5.25\,^{\circ}\simeq0.996$), but it is nevertheless included for accuracy. Hereafter, $\Delta l \cdot \cos b$ will be denoted $\Delta l$ for brevity.

\begin{figure}[ht!]
    \centering
    \includegraphics[width=0.95\linewidth]{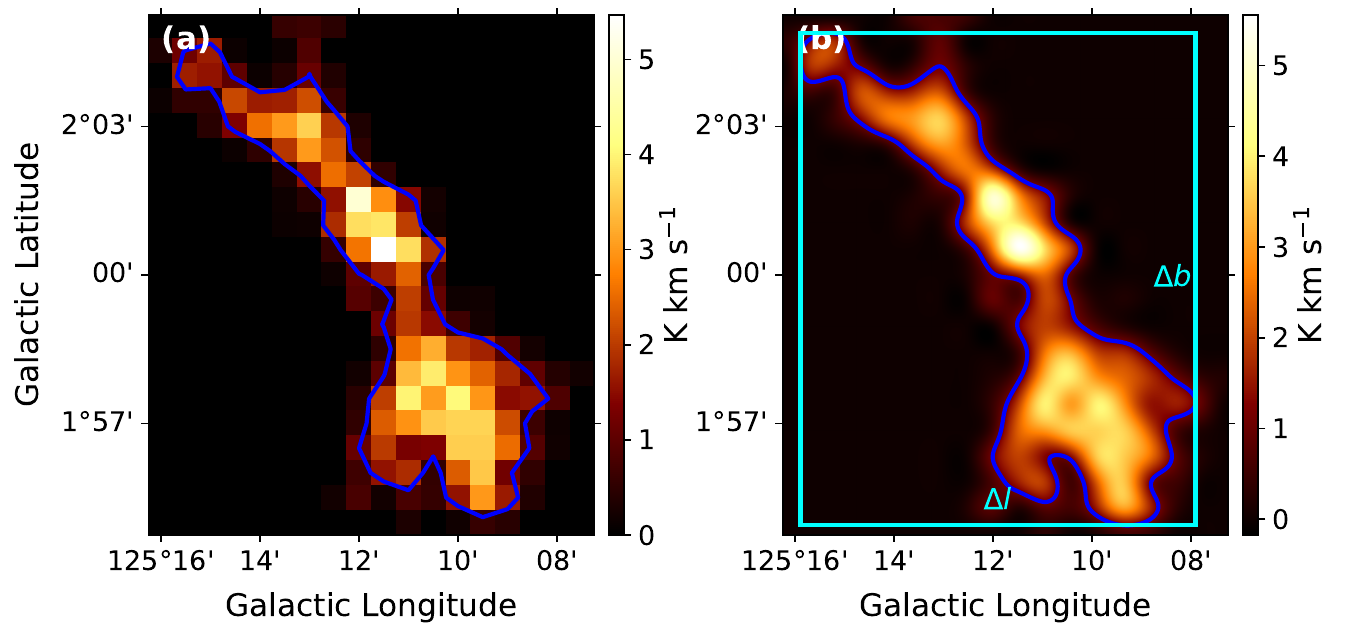}
    \includegraphics[width=0.95\linewidth]{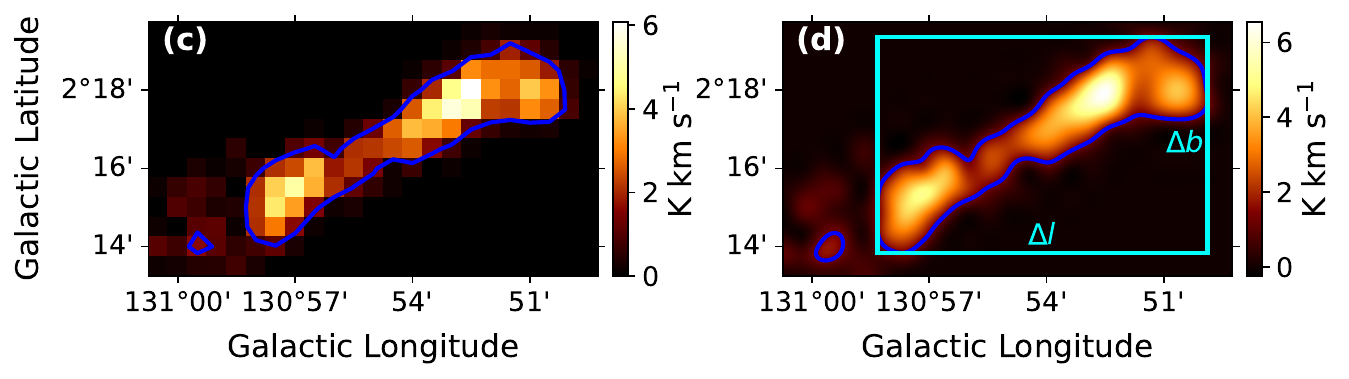}
    \caption{Example clouds structures from the sample. The left two panels show the original integrated-intensity maps, while the right two panels present the interpolated versions used for contour extraction. The contour corresponds to 20\% of the peak integrated-intensity. The cyan rectangles in the right two panels are bounding boxes fitted to the contours, with each width ($\Delta l$) and height ($\Delta b$) corresponding to the Galactic longitudinal and latitudinal directions, respectively. Panel (d) shows an example in which a tiny substructure is excluded from our analysis (bottom-left, area $< 1\,\mathrm{arcmin}^2$).}
    \label{fig:examples}
\end{figure}

To calculate the $\Delta b/\Delta l$ ratio, we derive $\Delta b$ and $\Delta l$ from the integrated-intensity contour of each molecular cloud. After obtaining the contour, we fit a rectangle aligned with Galactic longitude and latitude and define $\Delta l$ and $\Delta b$ as the lengths of the rectangle sides along the longitudinal and latitudinal directions, respectively (see Figure \ref{fig:examples}). To improve the precision of the $\Delta b/\Delta l$ ratio, we perform interpolation on the integrated-intensity map before the fitting.
We then use $\Delta b/\Delta l$ ratio to characterize whether a molecular cloud is more elongated along Galactic longitude or latitude.

Obviously the adopted thresholds for the boundary-defining contours will affect the fitted $\Delta l$ and $\Delta b$, and with a high threshold one large cloud may split into a few smaller ones.
We adopt 20\% of the peak integrated-intensity as a uniform threshold for substructure boundaries for all clouds.
Multiple connected contours of a large cloud are treated as multiple substructures.
After excluding regions with areas smaller than $1\,\mathrm{arcmin}^2$, the final sample consists of 20,022 substructures.

We estimate kinematic distances using the analytic Galactic rotation curve model of Russeil et al.\cite{Russeil+etal+2017}. For each source, the distance from the Sun is derived from its Galactic coordinates and $V_{\mathrm{LSR}}$, assuming a fixed solar motion and circular rotation.

We adopt a distance range of $0.5\text{--}6\,\mathrm{kpc}$. Below $0.5\,\mathrm{kpc}$, the velocity gradient is too weak, making the derived distances highly sensitive to peculiar motions. Beyond $6\,\mathrm{kpc}$, non-circular motions become increasingly important and the velocity–distance relation becomes less sensitive to distance, leading to larger uncertainties in the kinematic distance estimates. The adopted range therefore ensures the most reliable kinematic solutions for our sample.

\subsection{Structural, Kinematic, and Magnetic-Field Orientations}
\label{SubSec:Orientation_Method}

To characterize the orientation of each molecular cloud substructure, we fit ellipses to their integrated-intensity distributions. In this procedure, the integrated-intensity of each pixel is treated as the surface density of a rigid plate, and the best-fitting ellipse is derived from the moment of inertia of this plate about its center of mass (see Section~2.1 of Ge et al.\cite{Ge+etal+2024}). The resulting ellipse defines the orientation of the major axis in Galactic coordinates, which we adopt as the structural orientation.

The internal kinematics of each substructure are analyzed using the intensity-weighted mean velocity (Moment~1) map. Based on this map, we compute the two-dimensional velocity-gradient field and extract the distribution of gradient directions. Because the use of a central finite-difference scheme to compute gradients can introduce artificial spatial correlations when applied to interpolated velocity fields, we calculate the velocity-gradient field directly from the original velocity data without spatial interpolation.
We also note that the central finite-difference scheme itself introduces a certain level of spatial correlation, so we apply a dedicated procedure to mitigate this effect, as described in Appendix~\ref{Appendix:CentralDiff}.
These directional distributions are characterized by fitting a von~Mises function (see Appendix~\ref{Appendix:vonMises}), yielding both a principal gradient direction $\mu$ and a concentration parameter $\kappa$. Larger values of $\kappa$ indicate more coherently oriented velocity gradients and therefore more reliable direction estimates.
Plane-of-sky magnetic-field orientations are obtained from Planck $353\, \mathrm{GHz}$ polarization data, following the procedure described in Section~\ref{SubSec:Planck_353_GHz_Observation}.

To ensure robust comparisons between structural, kinematic, and magnetic-field orientations, we apply selection criteria based on the fitted velocity-gradient direction $\mu$ and the concentration parameter $\kappa$, as described in Appendix~\ref{Appendix:CentralDiff}. We further restrict the sample to substructures with distances in the range of $0.5\text{--}6\,\mathrm{kpc}$. After applying these criteria, a total of $644$ substructures remain in the final sample.

For the selected substructures, we compute the equivalent physical radius ($R_\mathrm{eq} = \sqrt{A / \pi}$, where $A$ is the projected physical area) and divide the sample into two subsamples with $R_\mathrm{eq} \le 2\,\mathrm{pc}$ ($N = 412$) and $R_\mathrm{eq} > 2\,\mathrm{pc}$ ($N = 232$), respectively, and perform the relative orientation analysis separately for these two subsamples.

\begin{figure}[ht!]
    \centering
    \includegraphics[width=0.6\linewidth]{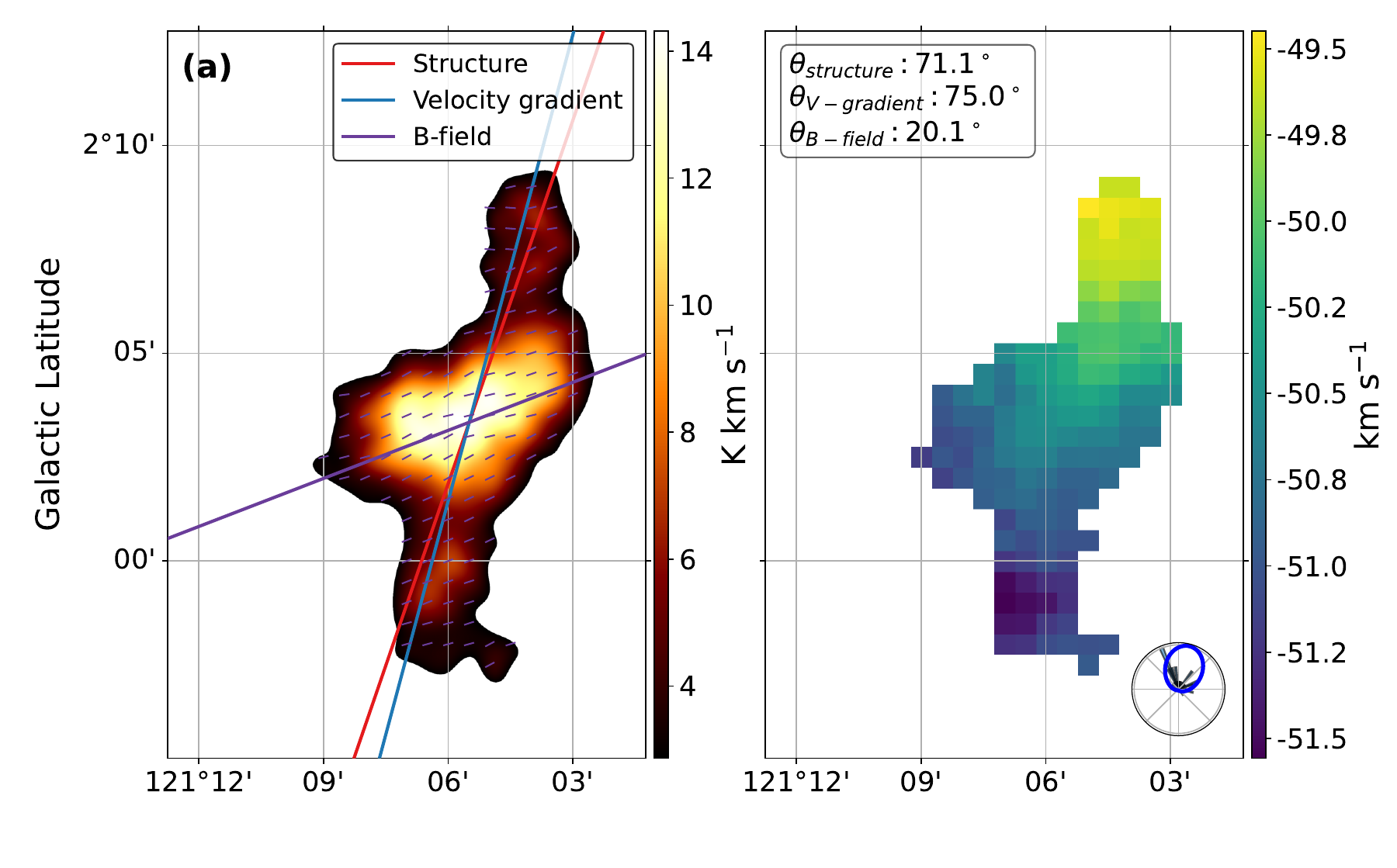}
    \includegraphics[width=0.6\linewidth]{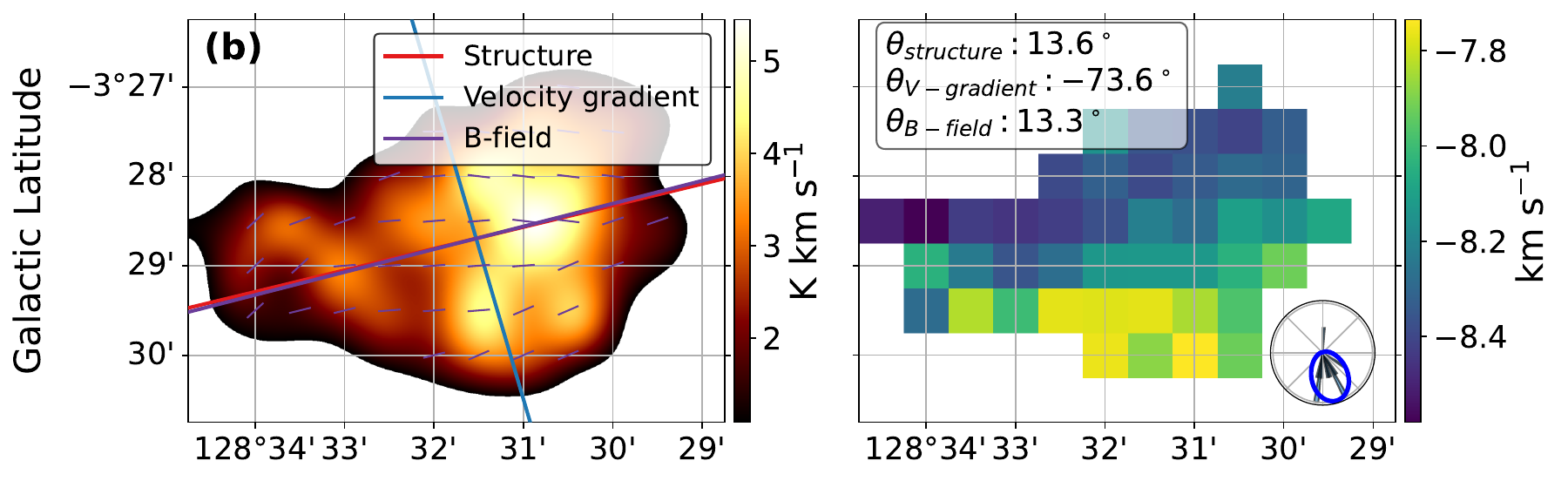}
    \includegraphics[width=0.6\linewidth]{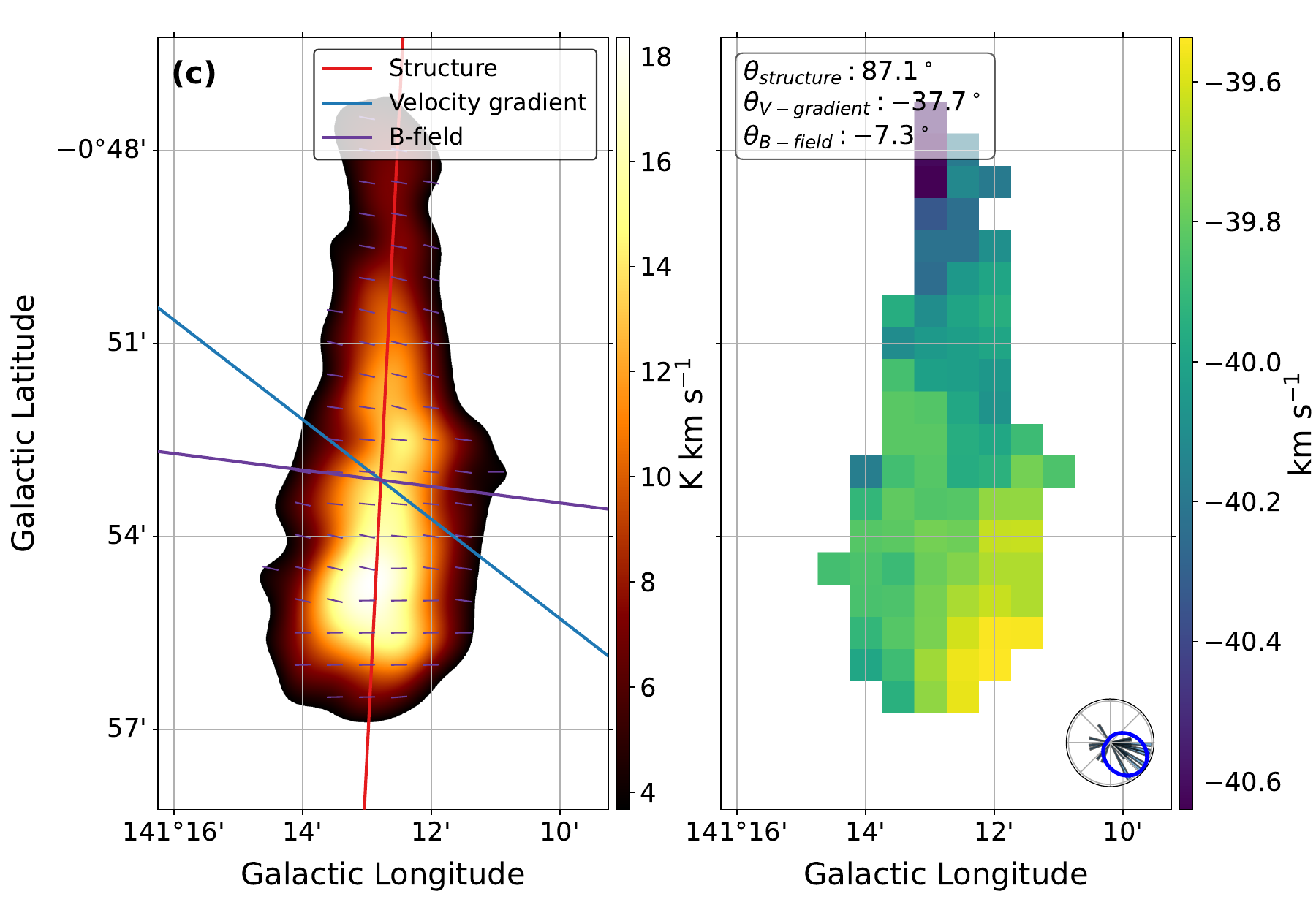}
    \caption{Three examples of substructures are shown: in panel (a), the structural orientation is parallel to the velocity-gradient direction; in panel (b), the structural orientation is perpendicular to the velocity-gradient direction and parallel to the magnetic-field direction; and in panel (c), the structural orientation is perpendicular to the magnetic-field direction. In each row, the left side displays the interpolated integrated-intensity map. Purple segments indicate the plane-of-sky magnetic-field directions derived from Planck data, the blue line marks the principal velocity-gradient direction obtained from von~Mises fits, and the red line indicates the structural orientation derived from ellipse fitting. The right side shows the velocity field, together with a circular diagram illustrating the von~Mises fitting results for the velocity-gradient directions.
    }
    \label{fig:12CO_masked_examples}
\end{figure}

Figure~\ref{fig:12CO_masked_examples} shows three representative examples of substructures, including their fitted orientations, velocity gradient directions, and magnetic field directions.

\section{Results and Discussion} \label{Sec:Results_and_Discussion}

\subsection{Distribution of \texorpdfstring{$\Delta b/\Delta l$}{Delta b/Delta l} Ratios in PPV Space} \label{Sec:Ratios_Distribution_in_PPV}

We divide the Galactic longitude into intervals of $5\,^{\circ}$ and the Galactic latitude into intervals of $1\,^{\circ}$. We also divide the velocity range into three intervals: $-95$ to $-60\,\mathrm{km\cdot s}^{-1}$, $-60$ to $-20\,\mathrm{km\cdot s}^{-1}$, and $-20$ to $25\,\mathrm{km\cdot s}^{-1}$. Based on the PPV (position-position-velocity) coordinates of each substructure, we assign its $\Delta b/\Delta l$ ratio to one of the resulting 270 ($=9\times10\times3$) bins.

\begin{figure}[ht!]
    \centering
    \includegraphics [width=0.95\linewidth]{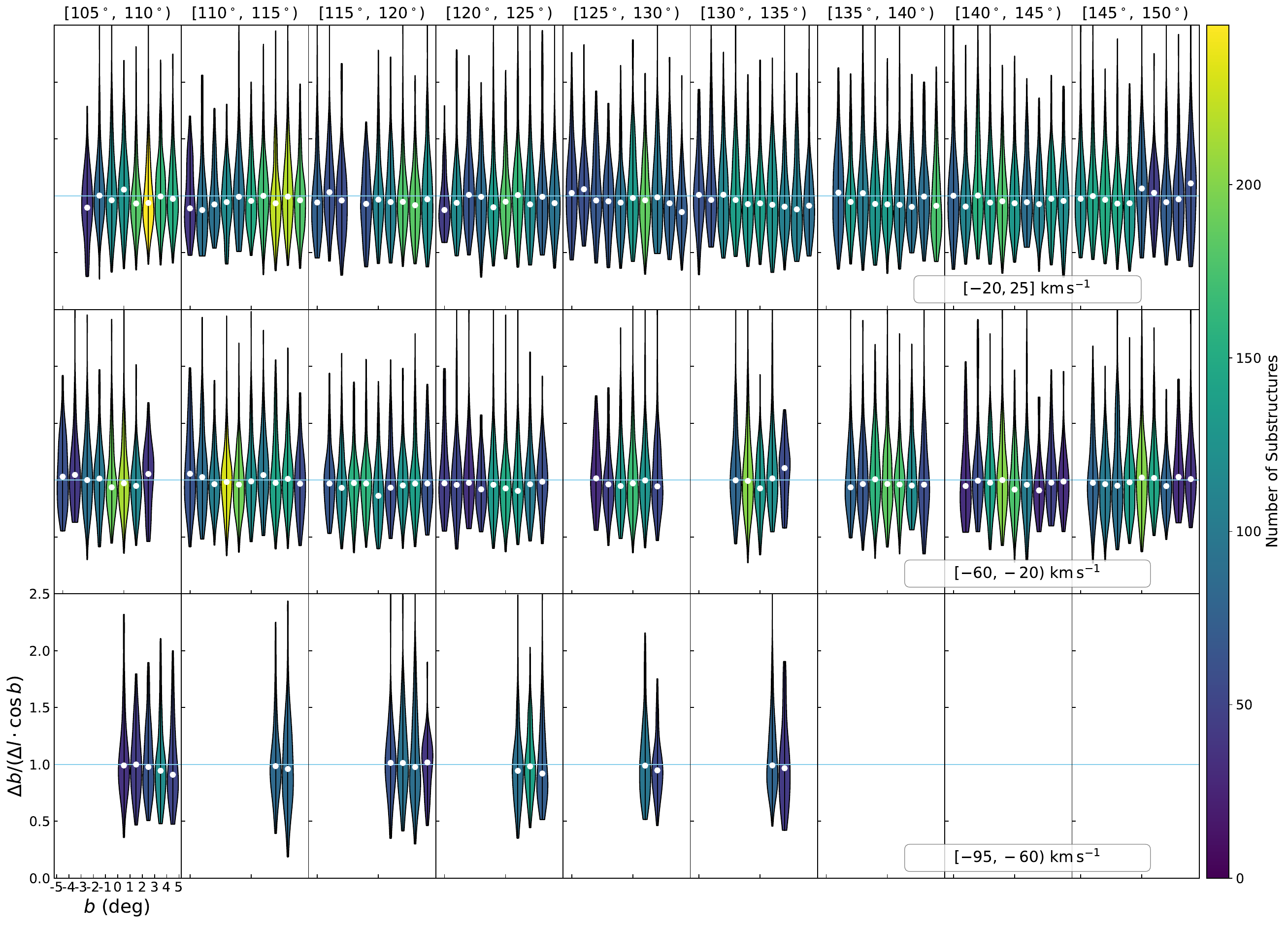}
    \caption{Violin plots showing the distribution of the $\Delta b/\Delta l$ ratio for molecular cloud substructures, grouped by Galactic longitude ($l$), Galactic latitude ($b$), and velocity ($v$). The $\Delta b/\Delta l$ ratios show no significant dependence on $l$, $b$, or $v$, but most are below 1 (the horizontal blue lines), indicating a slight but robust tendency that the substructures are preferentially elongated along the Galactic longitudinal direction.}
    \label{fig:violins_ar_substructures}
\end{figure}

Figure~\ref{fig:violins_ar_substructures} shows that, while the $\Delta b/\Delta l$ ratio exhibits no obvious dependence on Galactic latitude ($b$), longitude ($l$), or velocity ($v$), most substructures have $\Delta b/\Delta l < 1$, with a median value of 0.96, indicating a preference for elongation along Galactic longitude. A statistical resampling test of randomly generated Gaussian ellipses confirms that the observed median projected scale ratio is significantly different from unity (see Appendix~\ref{Appendix:Simulated_Gaussian_Ellipses} for details).

Our results are qualitatively consistent with Koda et al.\cite{Koda+etal+2006}. They analyzed 552 $^{13}$CO clouds in the inner Galaxy, covering the first Galactic quadrant, and measured their elongation axes, velocity gradients, and spin orientations. They found that the clouds are generally elongated along the Galactic plane, with axis-ratio peaks around 1.8.

\subsection{Correlation between \texorpdfstring{$\Delta b/\Delta l$}{Delta b/Delta l} Ratio and Substructure Size}

To investigate the possible physical origin of the result that the median value of the $\Delta b/\Delta l$ ratio is less than 1, we calculate two physical parameters related to substructure sizes and analyze their potential correlations with the $\Delta b/\Delta l$ ratio.

We first define an physical length to characterize each substructure: 
\begin{equation}
    L=d\cdot \sqrt{\Delta b^2 + (\Delta l \cdot \cos b)^2} \, ,
\end{equation}
where $d$ is the distance of the substructure.

\begin{figure}[ht!]
    \centering
    \includegraphics[width=0.95\linewidth]{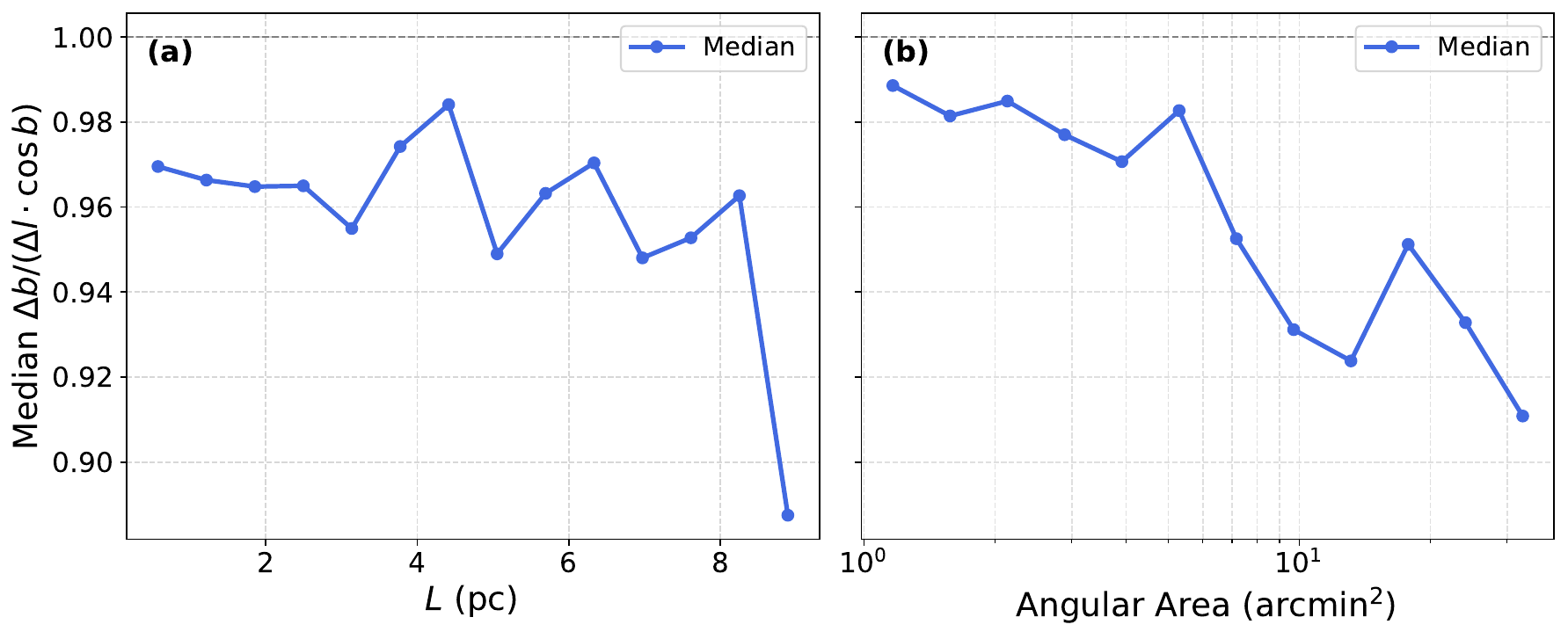}
    \caption{(a) Median $\Delta b/\Delta l$ ratio versus physical length $L_{\mathrm{eq}}$ (equal-width bins). (b) Median $\Delta b/(\Delta l\cdot\cos b)$ ratio versus angular area (logarithmic bins). The horizontal dashed line in both panels indicates $R=1$.}
    \label{fig:l_eq_and_area_vs_median_b_l}
\end{figure}

To further characterize the dependence of the $\Delta b/\Delta l$ ratio on substructure size, we bin the data in $L$ using equal-width intervals and compute the median $\Delta b/\Delta l$ ratio within each bin. Bins containing fewer than 200 substructures are excluded to ensure statistical robustness. The resulting trend shown in panel (a) of Figure~\ref{fig:l_eq_and_area_vs_median_b_l} is that the median $\Delta b/\Delta l$ ratio decreases systematically with increasing $L$, corroborating that larger substructures exhibit a stronger preference for elongation along Galactic longitude.
A similar trend is seen in the dependence of the $\Delta b/\Delta l$ ratio on angular area (panel (b) of Figure~\ref{fig:l_eq_and_area_vs_median_b_l}), where the data are binned in logarithmically spaced area intervals.

The behavior that larger substructures tend to become more flattened along the Galactic plane likely results from Galactic shear, which may enhance the longitudinal elongation of molecular structures, because differential rotation can stretch molecular material along the azimuthal direction. Similar processes have been demonstrated in hydrodynamic simulations of spiral galaxies, where differential rotation and divergent orbits in a spiral potential naturally shear dense gas structures into elongated features along the disk plane \cite{Dobbs+etal+2006}.
The effect of gravitational confinement, which pulls material towards the midplane and compresses molecular gas more strongly in the Galactic latitudinal ($b$) direction than in the longitudinal ($l$) direction, likely plays a minor role, as it primarily regulates the vertical scale height rather than producing systematic elongation within the Galactic plane.

\subsection{Relative Orientations of Structure, Velocity Gradients, and Magnetic Fields}
\label{SubSec:Relative_Orientations}

\begin{figure}[ht!]
    \centering
    \includegraphics[width=0.95\textwidth]{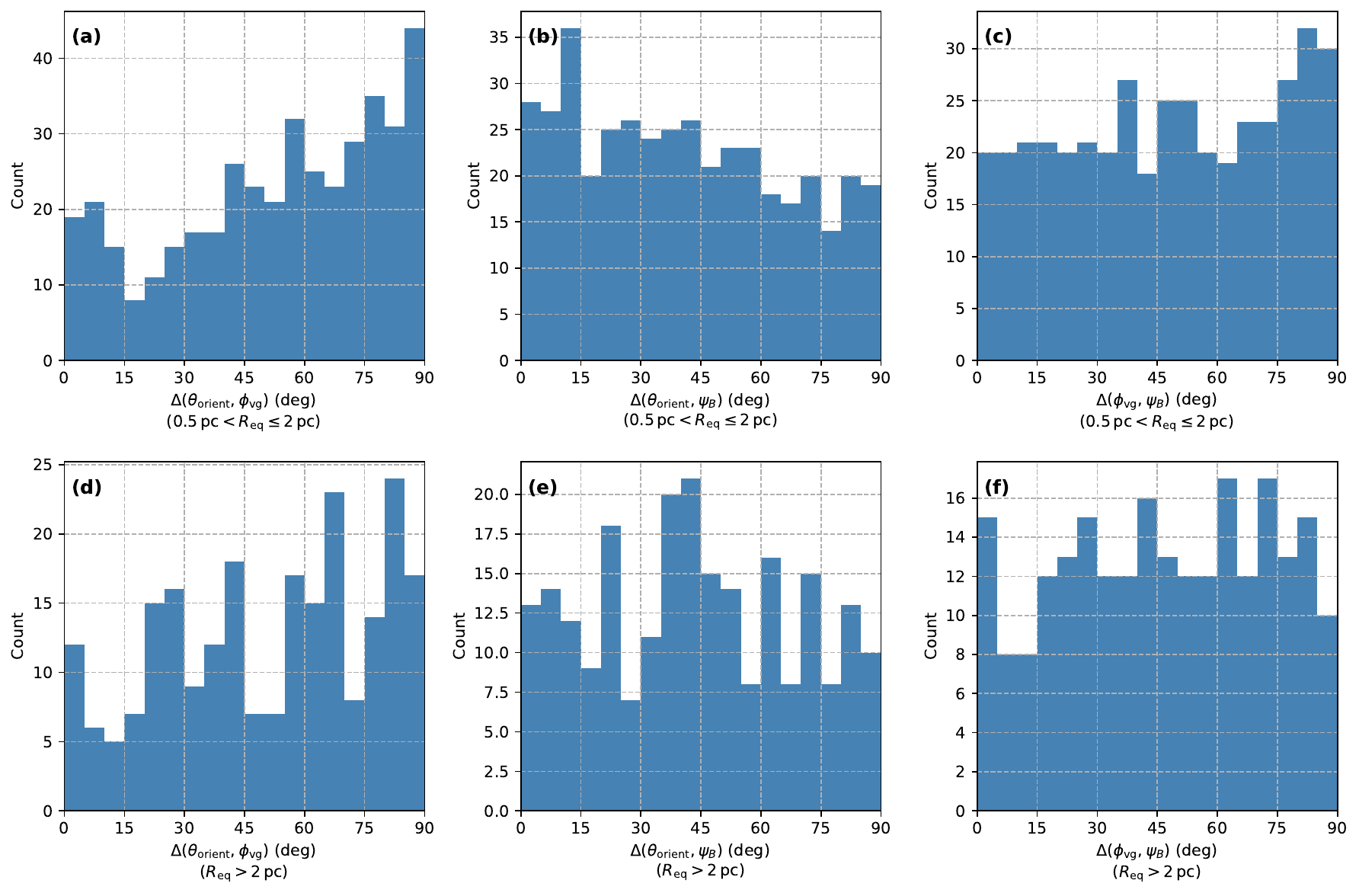}
    \caption{
    Distributions of relative angles between structural orientation, velocity-gradient direction, and plane-of-sky magnetic-field orientation. 
    Panels (a)-(c) correspond to substructures with $R_\mathrm{eq} \le 2\,\mathrm{pc}$, while panels (d)-(f) correspond to those with $R_\mathrm{eq} > 2\,\mathrm{pc}$. 
    The three columns show $\Delta(\theta_{\rm orient}, \phi_{\rm vg})$, $\Delta(\theta_{\rm orient}, \psi_B)$, and $\Delta(\phi_{\rm vg}, \psi_B)$, respectively. 
    All angles are folded into the range $0\,^\circ$-$90\,^\circ$.
    }
    \label{fig:relative_angle}
\end{figure}

For the selected sample of 644 substructures, we investigate the relative orientations among the structural major axes, velocity-gradient directions, and plane-of-sky magnetic-field orientations. The distributions of relative angles for two subsamples, defined by $R_\mathrm{eq} \le 2\,\mathrm{pc}$ and $R_\mathrm{eq} > 2\,\mathrm{pc}$, are shown in Figure~\ref{fig:relative_angle}.

It is worth noting that the results presented here are based on statistical orientation distributions and therefore reflect ensemble-level trends, rather than implying a direct causal relationship between magnetic fields and cloud dynamics.

For both subsamples, the distributions of $\Delta(\theta_{\rm orient}, \phi_{\rm vg})$ exhibit a pronounced peak around $90\,^\circ$, indicating that velocity gradients are preferentially perpendicular to the structural major axes. This systematic deviation from a uniform distribution suggests a non-random coupling between substructure morphology and internal kinematics. The distributions of $\Delta(\theta_{\mathrm{orient}}, \psi_B)$ show a scale-dependent behavior: for the smaller-scale subsample ($R_\mathrm{eq} \le 2\,\mathrm{pc}$) the distribution is skewed toward smaller angles (preferentially parallel), while for the larger-scale subsample ($R_\mathrm{eq} > 2\,\mathrm{pc}$) the distribution peaks near $40\,^\circ$, indicating no strong parallel or perpendicular preference. The relative orientations between velocity gradients and magnetic fields, $\Delta(\phi_{\mathrm{vg}}, \psi_B)$, show a scale-dependent trend: for $R_\mathrm{eq} \le 2\,\mathrm{pc}$ the distribution shows a slight peak near $90\,^{\circ}$ (weakly perpendicular), while for $R_\mathrm{eq} > 2\,\mathrm{pc}$ it is nearly uniform, indicating that the kinematic-magnetic alignment becomes less pronounced on larger scales.

Koda et al.\cite{Koda+etal+2006} reported that the directions of internal spin axes, defined as perpendicular to large-scale velocity gradients, are essentially random and uncorrelated with cloud elongation. In contrast, our results reveal a clear preference for velocity gradients to be perpendicular to the structural major axes in both subsamples, implying that any inferred spin axes would tend to align with the elongation direction. Nevertheless, this geometric configuration is not consistent with internal spin being the primary mechanism responsible for the observed elongation. Instead, it suggests a statistically non-random relationship between cloud morphology and internal kinematic properties.

Our finding of a non-random relationship between velocity-gradient directions and structural major axes may be qualitatively compared with results on much smaller and more evolved scales. Corsaro et al.\cite{Corsaro+etal+2017} reported spin-axis alignment among red giants in old open clusters (NGC 6791 and NGC 6819), suggesting that part of the initial angular momentum may be retained in stellar rotation over Gyr timescales. While the traced quantities differ, both studies are at least consistent with the possibility that non-random orientation patterns may exist across different evolutionary stages, from cloud kinematics to stellar rotation. However, this comparison is necessarily qualitative, as velocity gradients trace gas kinematics in molecular cloud substructures, whereas stellar spin axes reflect the integrated outcome of star formation and subsequent dynamical evolution, and a direct physical connection between the two therefore remains uncertain.

For comparison, Planck Collaboration et al.\cite{Planck+etal+2016} showed that plane-of-sky magnetic fields tend to align with structures in low-column-density environments and become increasingly perpendicular at higher column densities. The predominantly parallel alignment observed in the smaller-scale subsample is consistent with a low- to intermediate-column-density regime, while the larger-scale subsample shows no strong preferential orientation.

The scale dependence of the relative orientation between velocity gradients and magnetic fields has been highlighted in recent studies. Tang et al.\cite{Tang+etal+2019} found that on parsec scales velocity gradients can align with magnetic fields, whereas Chen et al.\cite{Chen+etal+2024} reported no preferred alignment on core scales ($\sim 0.05\,\mathrm{pc}$), suggesting that the observed kinematic–magnetic coupling is highly sensitive to the physical regime being probed.

Our results are consistent with this general picture but indicate a more complex transition. On smaller scales ($R_\mathrm{eq} \le 2\,\mathrm{pc}$), we find a weak preference for velocity gradients to be perpendicular to the magnetic field, while on larger scales ($R_\mathrm{eq} > 2\,\mathrm{pc}$) the distribution becomes nearly uniform. This suggests not only a weakening of alignment, but a change in the dominant physical processes governing the observed orientations.

Such behavior may reflect a scale-dependent interplay between multiple effects, rather than a simple weakening of magnetic influence. On larger scales, projection effects and Galactic differential rotation may contribute to the increasingly weak correspondence between velocity gradients and magnetic fields. Turbulent motions may further obscure any intrinsic alignment present on smaller scales.

In contrast, the correlation between cloud morphology and velocity gradients remains comparatively robust. This may indicate that velocity gradients may retain a closer correspondence with cloud morphology than with the magnetic field structure. Processes such as anisotropic collapse or convergent flows could naturally generate preferred orientations between cloud elongation and internal velocity structure. By comparison, a clear alignment between velocity gradients and magnetic fields is generally expected only when magnetic fields exert a sufficiently strong dynamical influence on turbulent motions.

It should also be noted that different observational studies probe different tracers, spatial scales, and dynamical environments, making direct comparisons difficult. The apparent differences between studies may therefore reflect complementary aspects of a multi-scale system rather than genuinely contradictory results.

\section{Conclusions} \label{Sec:Conclusions}

We analyzed the morphology and orientation of molecular cloud substructures in the MWISP $^{12}\mathrm{CO}\,(1\text{--}0)$ survey using the observational projected scale ratio $R = \Delta b / (\Delta l \cdot \cos b)$, derived from integrated-intensity contours with strict quality control. The median $R < 1$ indicates a statistical tendency for elongation along Galactic longitude, which becomes more pronounced for larger substructures, whether defined by physical length or angular area. Monte Carlo simulations of two-dimensional Gaussian ellipses yield a median $\Delta b/\Delta l$ ratio near unity, confirming that the observed anisotropy cannot be attributed solely to measurement geometry.

An analysis of relative orientations among structural major axes, velocity-gradient directions, and Planck $353\, \mathrm{GHz}$ magnetic-field angles suggests differences between the two scale regimes in the relationship between morphology, kinematics, and magnetic fields. Velocity gradients are preferentially perpendicular to the structural major axes in both subsamples, indicating a systematic correlation between morphology and internal kinematics. 

The alignment between velocity gradients and magnetic fields becomes less distinct toward larger scales, transitioning from a weakly perpendicular tendency at $R_\mathrm{eq} \le 2\,\mathrm{pc}$ to a distribution consistent with a more random orientation at larger scales. The orientations between structure and magnetic fields are generally parallel or weakly misaligned, consistent with a low- to intermediate-column-density regime.

These results suggest that molecular cloud anisotropy may arise from a combination of processes, including large-scale Galactic dynamics, anisotropic gas flows, and magnetic fields. The comparatively stronger correlation between morphology and velocity gradients indicates that cloud kinematics may retain a closer connection to structural evolution than to magnetic-field orientation on the scales probed here. The weaker velocity gradient--magnetic field alignment at larger scales may reflect the increasing influence of projection effects, turbulent motions, and Galactic differential rotation, which can obscure intrinsic kinematic--magnetic coupling.

Since different observational studies probe different tracers, spatial scales, and dynamical regimes, direct comparisons between results are not always straightforward. Our findings therefore provide a complementary view of the scale-dependent interplay between morphology, kinematics, and magnetic fields in molecular clouds, and offer observational constraints on the multi-scale physics of cloud formation and evolution.







\newpage
\appendix

\section{Correction of Artificial Correlations in Gradient Fields Computed with a Central Finite-Difference Scheme}
\label{Appendix:CentralDiff}

\setcounter{figure}{0}
\setcounter{equation}{0}
\renewcommand{\thefigure}{A1-\arabic{figure}}
\renewcommand{\theequation}{A1-\arabic{equation}}

When computing velocity gradients using central differences within finite rectangular regions, the geometry of the region can introduce an artificial bias in the estimated mean gradient direction obtained via the vector-average method (MLE). To illustrate this effect, consider a 2D rectangular mask of size $(N_x+2) \times (N_y+2)$ containing a completely random velocity field. Ideally, the gradient field should exhibit no preferred direction. For sufficiently large regions, we neglect the gradients at the outermost pixel layer and compute velocity gradients at interior points using central differences along the $x$ and $y$ directions:
\begin{equation}
    g_{x,ij} = \frac{v_{i+1,j} - v_{i-1,j}}{2} \, , \quad
    g_{y,ij} = \frac{v_{i,j+1} - v_{i,j-1}}{2} \, ,
\end{equation}
we sum the gradients along each direction to compute the mean gradient vector:
\begin{equation}
    \vec{G}_{\mathrm{mean}}
    =
    \left(
    \frac{1}{N_y}\sum_{j=1}^{N_y}
    \left[
    \frac{1}{N_x}\sum_{i=1}^{N_x} g_{x,ij}
    \right] \, ,
    \;
    \frac{1}{N_x}\sum_{i=1}^{N_x}
    \left[
    \frac{1}{N_y}\sum_{j=1}^{N_y} g_{y,ij}
    \right]
    \right) \, .
\end{equation}

Without loss of generality, we consider the case $N_x \gg N_y$, so that the $x$-direction corresponds to the long edge of the region. For the $x$-direction (long edge) accumulation, interior terms largely cancel due to the telescoping property of central differences:
\begin{equation}
    \sum_{i=1}^{N_x} g_{x,ij} = \frac{1}{2} \big[ (v_{N_x+1,j} - v_{N_x,j}) + (v_{1,j} - v_{0,j}) \big] \, ,
\end{equation}
so only the contributions from the boundary points remain. Since the number of points along the long edge $N_x$ is large, the fraction of points contributing to the residual vector is small. Conversely, for the short edge ($y$-direction) with $N_y \ll N_x$, boundary contributions constitute a larger fraction, leading to a stronger residual along $y$.  

The additional averaging over $j$ in the outer summation does not alter this conclusion, because $g_{x,ij}$ values at different $j$ are statistically uncorrelated in a random field. An analogous argument applies to the $y$-direction, where gradients at different $i$ are likewise uncorrelated.

Consequently, the vector-average MLE is dominated by contributions from the short edge, producing an apparent mean gradient orientation
\begin{equation}
    \mu
    =
    \arctan\left(
    G_{\mathrm{mean},y}/
    G_{\mathrm{mean},x}
    \right)
    \approx
    \text{along the short edge (perpendicular to the long axis)} \, .
\end{equation}

\begin{figure}[ht!]
    \centering
    \includegraphics [width=0.95\linewidth]{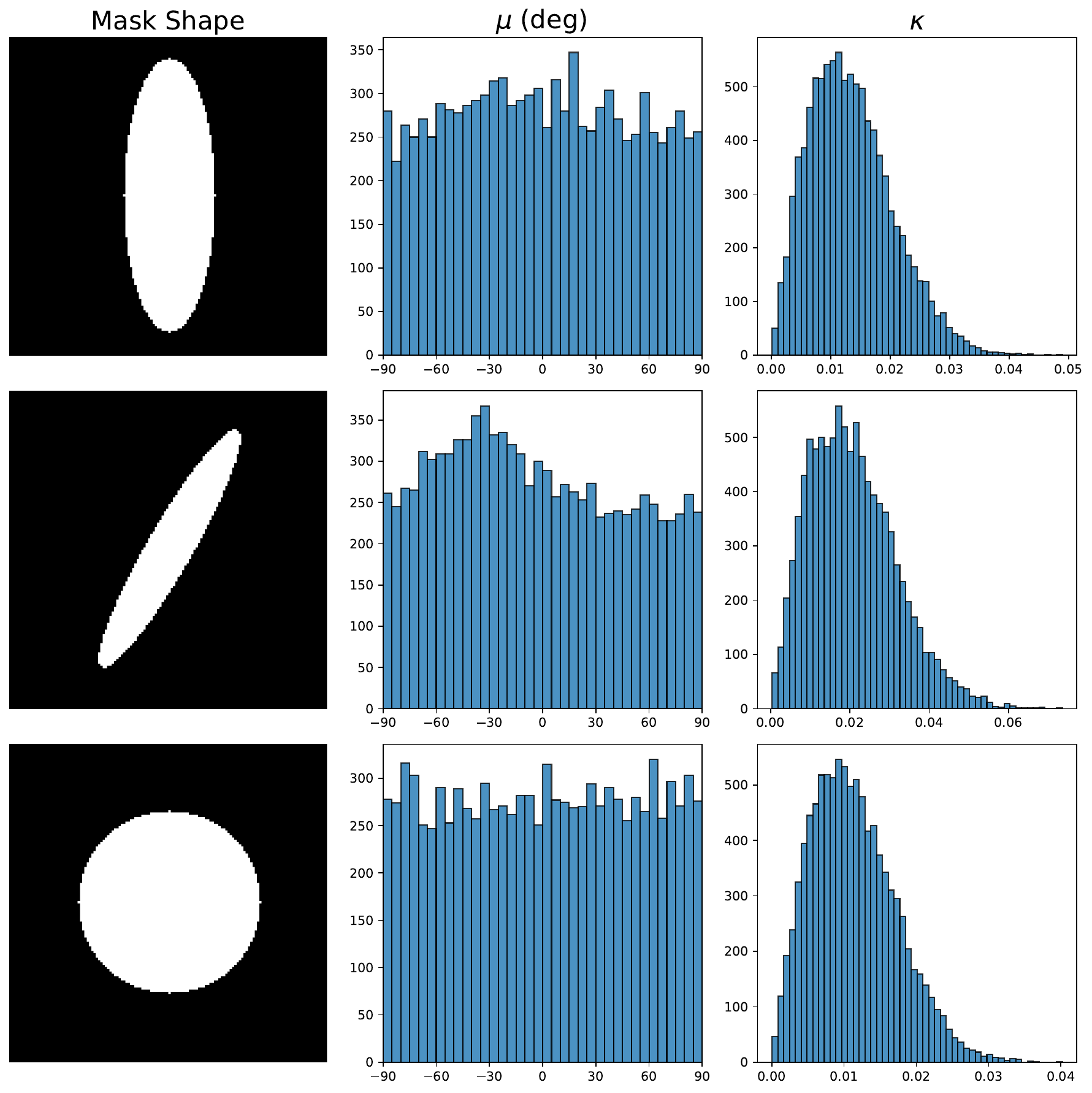}
    \caption{Monte Carlo results for velocity gradients computed in elliptical regions of different projected scale ratios. The first column shows the geometry of the sampling masks. The second column displays the distributions of the fitted mean gradient direction $\mu$, illustrating the artificial alignment induced by elongated masks. The third column shows the distributions of the concentration parameter $\kappa$, which are generally low, indicating weak directional coherence in the random fields. These results demonstrate the geometric bias introduced by central finite-difference gradients when combined with vector-averaging estimators.}
    \label{fig:gradient_test}
\end{figure}

To quantify this effect, we conduct Monte Carlo experiments in which velocity fields are generated from a uniform random distribution and sampled within regions of different geometries. The resulting distributions of the fitted parameters $\mu$ and $\kappa$ show that elongated domains systematically yield nonzero resultant vectors, consistent with the predicted geometric bias (see Figure~\ref{fig:gradient_test}).

The principal directions obtained from the simulations correspond to very low values of $\kappa$, so filtering based on $\kappa$ can partially mitigate this effect. However, we still recommend applying the following procedure to fully eliminate it.

\begin{figure}[ht!]
    \centering
    \includegraphics [width=0.95\linewidth]{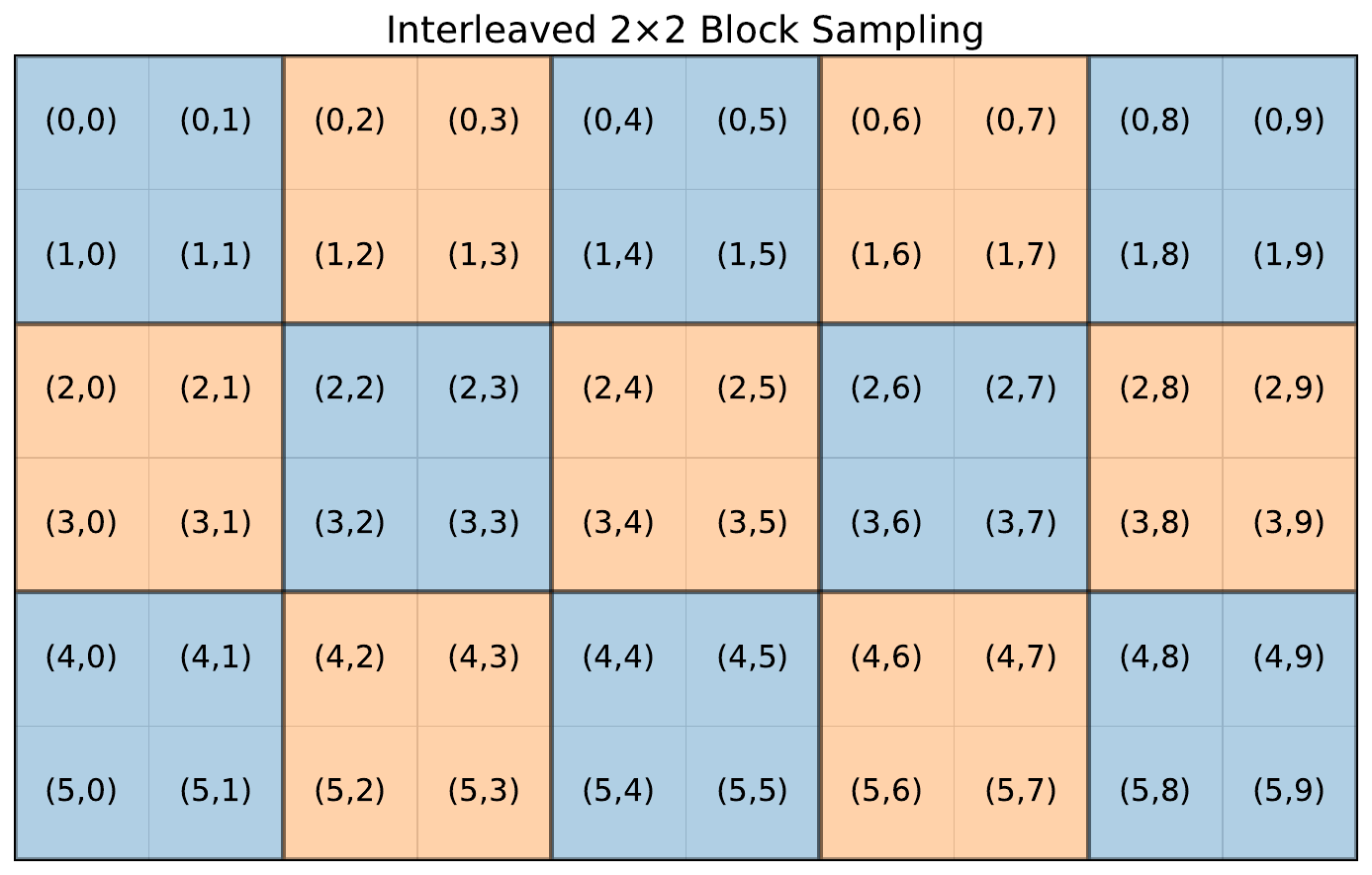}
    \caption{Schematic illustration of the interleaved $2\times2$ block-sampling scheme used to suppress artificial spatial correlations in gradient fields computed via central differences. Individual pixels are labeled by their array indices $(i,j)$ and grouped into non-overlapping $2\times2$ blocks. Blocks are selected in an alternating checkerboard pattern to construct two statistically independent subsamples, thereby reducing geometry-driven biases in subsequent vector-averaging analyses.}
    \label{fig:interleaved_block_sampling}
\end{figure}

As illustrated in Figure~\ref{fig:interleaved_block_sampling}, each pixel retains its original array index, while the block selection alternates between adjacent rows and columns. This construction effectively breaks the pixel-to-pixel correlations induced by central differencing. The resulting subsamples can therefore be used for independent gradient estimation and subsequent statistical analysis without introducing geometry-driven biases.

\begin{figure}[ht!]
    \centering
    \includegraphics [width=0.95\linewidth]{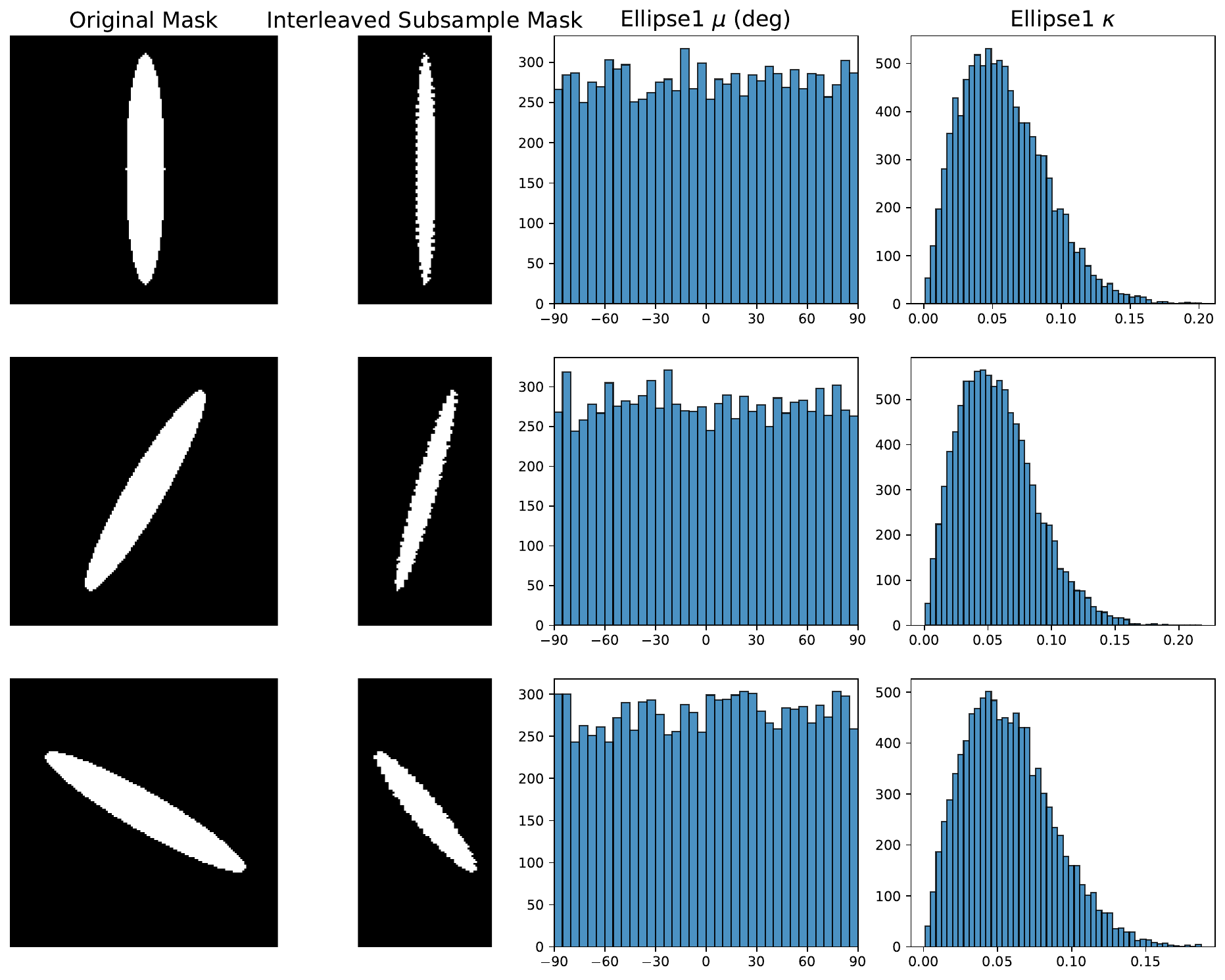}
    \caption{Monte Carlo results using the interleaved block-sampling scheme illustrated in Figure~\ref{fig:interleaved_block_sampling}. The first column shows the original rectangular masks. The second column displays the interleaved block-sampled masks used to break pixel-to-pixel correlations. The third column shows the distributions of the fitted mean gradient direction $\mu$, which are now consistent with a uniform distribution, indicating that the artificial geometric bias has been effectively removed. The fourth column shows the distributions of the concentration parameter $\kappa$, which remain low, reflecting weak directional coherence in the random velocity fields.}
    \label{fig:gradient_test2}
\end{figure}

We repeat the Monte Carlo experiments using this interleaved block-sampling scheme. As shown in Figure~\ref{fig:gradient_test2}, the distributions of $\mu$ are now uniform, confirming that the artificial bias has been effectively eliminated.

The excluded pixels are statistically equivalent to the selected ones and can be combined to form an alternative subsample, which allows an independent von~Mises fit (see Appendix~\ref{Appendix:vonMises}) as well. In practical applications to observational data, this procedure yields two principal gradient directions for a given substructure. We regard the gradient orientation as reliable when the angular difference between the two directions is smaller than $15\,^\circ$ and both fits yield concentration parameters $\kappa > 1$.

\section{Von~Mises Distribution Fitting of Velocity Gradient Directions}
\label{Appendix:vonMises}

\setcounter{equation}{0}
\renewcommand{\theequation}{A2-\arabic{equation}}

The von~Mises distribution is a circular probability distribution suitable for modeling angles or directional quantities. Its probability density function is

\begin{equation}
f(\theta \,|\, \mu, \kappa) = \frac{1}{2\pi I_0(\kappa)} \exp\big(\kappa \cos(\theta - \mu)\big) \, ,
\end{equation}

where $\theta$ is the angular variable, $I_0(\kappa)$ is the modified Bessel function of the first kind of order zero, $\mu \in [0, 2\pi)$ denotes the mean direction, and $\kappa \ge 0$ measures the concentration of the distribution around $\mu$.

For each molecular cloud substructure, we fit a von~Mises distribution to the measured velocity gradient angles. The fitting is performed using maximum likelihood estimation (MLE), which produces point estimates of the mean direction ($\mu$) and the concentration parameter ($\kappa$). The mean direction $\mu$ represents the preferred orientation of the velocity gradients, while the concentration $\kappa$ quantifies the directional coherence, with larger values indicating that the angles are more tightly clustered around $\mu$.

It is important to note that the estimated parameters $\mu$ and $\kappa$ are point estimates derived from the observed data and characterize the underlying angular distribution of the measured velocity gradients. To ensure reliable characterization, we retain substructures for further analysis only if they meet predefined thresholds in both $\kappa$ and the angular coverage of valid gradient measurements.

\section{Control Test with Simulated Gaussian Ellipses} 
\label{Appendix:Simulated_Gaussian_Ellipses}

\setcounter{figure}{0}
\renewcommand{\thefigure}{A3-\arabic{figure}}

We perform a control test using simulated Gaussian ellipses to assess whether the observed $\Delta b/\Delta l$ anisotropy could arise from measurement effects or nuances in the data handling alone. In this test, we analyze the $\Delta y/\Delta x$ ratios of simulated structures, which directly correspond to the $\Delta b/\Delta l$ ratios defined for the observational data.

In the simulations, we generate two-dimensional Gaussian elliptical distributions by randomly sampling the standard deviations of the major and minor axes ($\sigma_x$, $\sigma_y$) and the rotation angle within prescribed ranges. Following the $2\sigma$ cutoff adopted for the MWISP data, we apply an intensity threshold by setting values below the threshold to zero. All subsequent procedures are performed in exactly the same manner as described in Section~\ref{SubSec:Selection_and_PSR_Definition}.

\begin{figure}[ht!]
    \centering
    \includegraphics [width=0.95\linewidth]{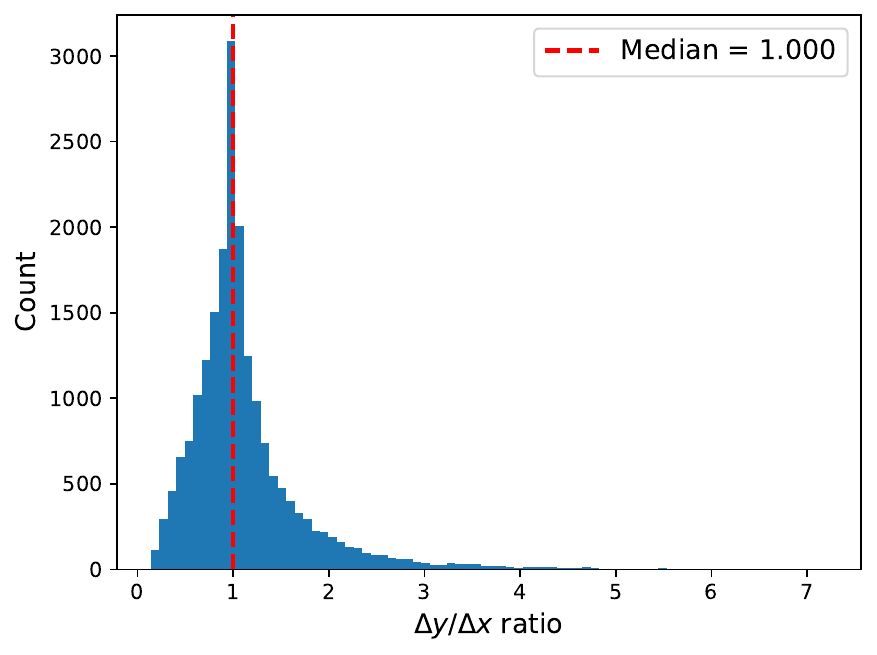}
    \caption{The $\Delta y/\Delta x$ ratio distribution of the simulated substructures.}
    \label{fig:simulation_result}
\end{figure}

Figure~\ref{fig:simulation_result} presents the $\Delta y/\Delta x$ ratio distribution of the simulated substructures, with vertical dashed lines marking the median value. The resulting median ratio is close to unity, in clear contrast to the observational result. This comparison indicates that the anisotropy observed in real molecular clouds is unlikely to be produced by measurement geometry or analysis procedures alone, and instead likely reflects additional physical influences.

\end{document}